\documentclass{statsoc}
\usepackage{natbib}
\usepackage{graphicx}
\usepackage{multirow}
\usepackage[centertags]{amsmath}
\usepackage[letterpaper]{geometry}
\newcommand{\var}{\mbox{Var}}

\newcommand{\E}{\mbox{E}}
\newcommand{\X}{\mathbf{X}}
\newcommand{\V}{\mathbf{V}}
\newcommand{\Z}{\mathbf{Z}}
\newcommand{\y}{\mathbf{y}}
\newcommand{\ub}{\mathbf{u}}
\newcommand{\T}{\scriptscriptstyle \mathrm{T}}
\newcommand{\uno}{\mathbf{1}}

\DeclareGraphicsExtensions{.eps,.pdf,.png,.jpg}

\title[A Multilevel Model with Autoregressive Components]{The Analysis of Tribal Art Prices:\\ a Multilevel Model with Autoregressive Components}
\author[Modugno L.]{Lucia Modugno}
\address{Department of Statistical Sciences, University of Bologna, Italy.}
\email{lucia.modugno@unibo.it}
\author[Cagnone S.]{Silvia Cagnone\footnote{\emph{Address for correspondence}: Silvia Cagnone, Department of Statistical Sciences, University of Bologna, Via Belle Arti, 41 - 40126 Bologna, Italy. E-mail: silvia.cagnone@unibo.it.}}\address{Department of Statistical Sciences, University of Bologna, Italy.}
\email{silvia.cagnone@unibo.it}
\author[L.Modugno S.Giannerini S.Cagnone]{Simone Giannerini}
\address{Department of Statistical Sciences, University of Bologna, Italy.}
\email{simone.giannerini@unibo.it}


\begin{document}
\begin{abstract}
    In this paper we propose a multilevel model specification with time series components for the analysis of prices of artworks sold at auctions. Since auction data do not constitute a panel or a time series but are composed of repeated cross-sections they require a specification with items at the first level nested in time points. An original feature of our approach is the derivation of full maximum likelihood estimators through the E-M algorithm. The data analysed come from the first database of ethnic artworks sold in the most important auctions worldwide. The results show that the new specification improves considerably over existing proposals both in terms of fit and prediction.
  \end{abstract}

\section{Introduction}\label{sec:intro}
Nowadays, artwork items are considered investment assets similarly to stocks, bonds and real estates. For this reason, in the recent past, the analysis of this new market segment was performed by resorting to tools for the analysis of financial markets. However, such tools miss some essential aspects of the art market. Indeed, contrarily to stocks that are exchanged a high number of times in each instant, artworks are one-off pieces of their kind, hardly comparable with each other, and they pass through the market only a handful of times (usually only one). A further substantial difference with respect to financial assets is that works of art provide aesthetic pleasure and social status to its owner other than mere monetary returns \citep{Goetzmann}. Moreover, there are considerable transaction costs and, last but not least, there are no publicly available good databases on this segment. Hence, the study of the art market requires new tools and renovated research efforts.
\par
One of the most important problems in the analysis of art markets is the study of price indexes for artwork items. In the Art Economics literature several proposals have been discussed, especially for paintings. Among the most important contributions we mention \emph{Sotheby's Art Index} (and similar others), the \emph{average painting methodology}~\citep{Stein}, the \emph{representative painting method}~\citep{Candela}, the \emph{repeated sales regression}~\citep{Goetzmann} and the \emph{hedonic regression}, called also the \emph{grey painting method}. The hedonic regression model is the most used approach for modelling art prices; the idea is due to~\citet{Rosen}, whereas developments and applications can be found in~\citet{Chanel,Ginsburgh,Agnello,Chanel2,Collins,Locatelli}. The method assumes that the price of an artwork depends both on the market trend and on certain object characteristics. Such dependence is modelled through a fixed effect regression. In particular, the estimated regression coefficients are interpreted as the price of each feature, the so-called \emph{shadow price}, assumed to be constant over time. Hence, it is possible to predict the price of a given object by summing the prices of its features. Also, a time-dependent intercept can represent the value of the \emph{grey painting} in that period, that is, the value of an artwork created by a standard artist, through standard techniques, with standard dimensions, etc.~\citep{Candela04,Locatelli}. The final market price index is built from the prices of the \emph{grey painting} in different periods. The hedonic regression model has the advantage of solving the problem of artwork heterogeneity by explaining prices through object features; also, it allows to derive a price index by neutralizing the effect of quality. Nevertheless, such method presents several drawbacks. First of all, it is difficult to account for all the relevant features that determine the price of an object, so that only a part of the price is explained. Moreover, most of the object features are categorical, such as, for example, the artist's name that in Western art strongly affects the price of artworks. Therefore, the regression equation will contain many dummy variables and, consequently, a high number of parameters to be estimated, so that the resulting models are not parsimonious. Most importantly, it is not possible to forecast prices as the time dynamics is not modelled explicitly. In fact the price index relies only on the estimated coefficients of time-dependent dummy variables.
\par
In order to overcome the limits of the hedonic regression model, we propose a multilevel approach. Multilevel data~\citep{Goldstein2010,Laird,Raudenbush} consist of units of analysis of different type, one hierarchically clustered within the other. At the lowest level (level-1 observations) such units can be described by some variables; furthermore, they are also grouped into larger units (higher level observations), which in turn could be described by other variables. The general specification of multilevel models~\citep{Skrondal_book} allows a large variety of applications. In particular, repeated measures data can be seen as a specific case of multilevel data with occasions $i$ at level-1 and units $j$ at level 2~\citep{Leeden98,Maas03}. The dependence among level-1 errors that characterize panel data can be handled by including correlation structures at level-1~\citep{Goldstein2010}. For instance, \citet{Jones93} and~\citet{Vonesh} model the residual errors through a first-order autoregression or
autoregressive moving-average (ARMA) processes. Moreover, it is possible to allow heteroscedastic within-group errors through variance functions~\citep{Davidian}. This flexibility in the specification of covariance structures represents an important feature of linear mixed-effect models for longitudinal data. In all these cases, any time dependence is modelled at the first level.
\par
Since auction data do not constitute a proper panel the multilevel approach for longitudinal data described above cannot be applied. Indeed, auction data have a structure similar to that of repeated cross-sectional surveys. The main aim of this work is to propose a multilevel model specification that is particularly suitable to handling prices of artworks sold at auctions over time. Such data consist of observations on individual survey respondents drawn from the same context (e.g. the same country) at many different time-points; therefore, they can be clustered in time-points \citep{Firebaugh} so that, even if it is not possible to follow specific individuals over time, they allow to catch social changes. \citet{DiPrete} were the first to adopt a multilevel framework to analyze repeated cross-sectional data. They called their model \emph{single-context multilevel model} as opposed to the traditional \emph{multiple-contexts} model. The substantial difference with traditional models is the serial correlation among level-2 units/time-points. The authors took into account this case by deriving a generalized least-square estimator.  A similar idea has been considered by \citet{Browne} but for spatial correlations and in a Bayesian framework. In their work, the independence assumption among level-2 disturbances is relaxed and the correlation between pairs of clusters is modelled through an explicit function of the distance between them. However, to our knowledge, such approaches are not implemented in any software handling multilevel model. Moreover, a full maximum likelihood approach for this specification has not been considered. For these reasons, despite the wide potential interest, multilevel models for this kind of data are poorly developed and seldom applied.
\par
In this paper we aim to fill in these gaps in many respects. We derive full maximum likelihood estimators with known desirable properties for the multilevel specification similar to that presented in \citet{DiPrete}. We treat auction data as repeated cross-sections by taking individuals (in our case artwork items) as level-1 units and time-points as level-2 units. Hence, the price dynamics over time are modelled at the second level by means of an autoregressive structure of first order between random effects, as required by the case under investigation. The proposal combines the flexibility of mixed effect models together with the predicting performance of time series components. This specification turns out to be a natural and more convenient choice over the hedonic regression for modelling artwork prices. The overall result is a parsimonious yet powerful specification that can also reveal a useful tool to forecast the future values of the price. We obtain model estimates through an EM iterative algorithm and derive robust standard errors by means of a bootstrap scheme. The work has been motivated by the analysis of the first world database of Tribal art prices. Such database has been built by a team of researchers of the University of Bologna, Faculty of Economics -- Rimini, in conjunction with other institutions, and contains information on over 20000 artwork items sold by the most important auction houses from 1998 onwards.
\par
The paper is organized as follows. The database of Tribal art prices is described in the next Section. In Section~\ref{application} we present the multilevel specification for Tribal art data and compare it with the traditional hedonic approach. Section~\ref{newmodel} contains the main theoretical contribution of this paper, that is the extension of the multilevel model to deal with the time dependence at the second level through a maximum likelihood approach. Section~\ref{app} describes the results of the new model fitted on Tribal art data and compares them with those of the classic version. Also, the predictive performance of the three models is assessed. Finally, conclusions and discussions of future research are provided in Section~\ref{sec:concl}.

\section{The first database of Tribal artworks}\label{sec:data}
\begin{figure}
    \centering
    \includegraphics[width=6cm,height=6cm]{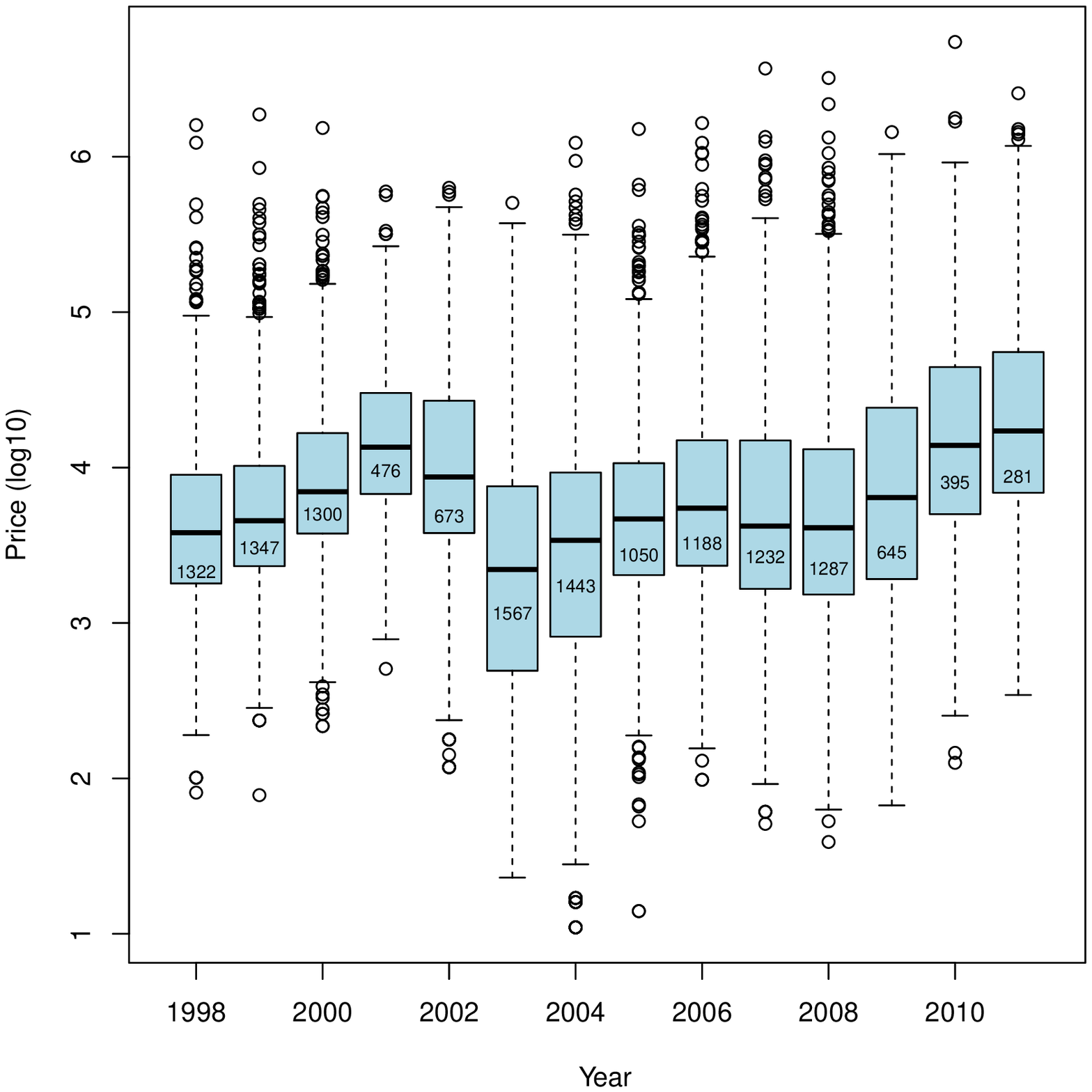}
    \includegraphics[width=6cm,height=6cm]{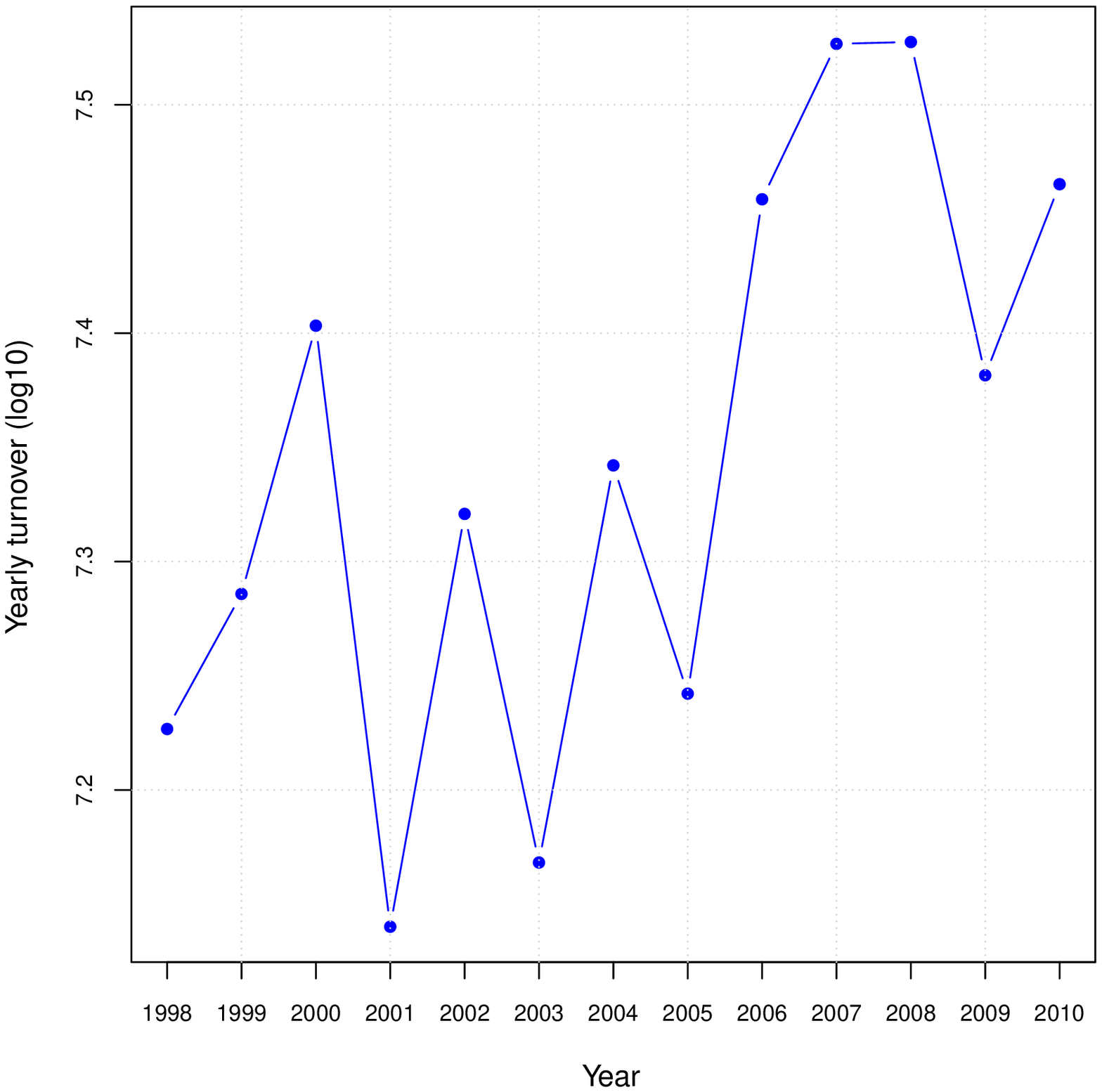}
    \caption{(left): yearly turnover (euro) in logarithmic scale (base 10) of the Tribal art market; (right): boxplots of prices by year. The amount of items sold in a given year is reported inside the boxes.}\label{fig:series}
\end{figure}

The important problem of the construction of price indexes for art markets requires data on sales of artworks. At present, the only available information in this area comes from auction exchanges. Nowadays, there are private companies (e.g. Artnet.com, Artinfo.com, Arsvalue.com, Artprice.com) that publish and sell information about auctions and price indexes, as well as art evaluations and other services. However, most of these companies deal with Western art. In this scenario, for a long time, there has not been a database on the Ethnic art. In recent years, the turnover of the Tribal art market (see Figure~\ref{fig:series}(left)) attracted the interest of investors and economists.  The first database on Ethnic artworks has been created in 2006 from the agreement of four institutions: the Department of Economics of the University of the Italian Switzerland, the Museum of the Extraeuropean cultures in Lugano, the Museo degli Sguardi in Rimini, and the Faculty of Economics of the University of Bologna, campus of Rimini. For each object, 37 variables are recorded from the paper catalogues released by the auction houses before the auctions; such variables include physical, historical and market characteristics. After the auction, the information on the selling price is added to the record.

In Figure~\ref{fig:series}(left) we report the boxplots of logged prices aggregated by year (inside the boxes, the total amount of item sold in each year). The plot provides a visual description of the structure of the dataset: a different group of artworks is sold each year; e.g. 1322 items were auctioned in 1998, 1347 objects different from the first set are sold in 1999, and so on. It is clear that Tribal art data do not constitute either a panel or a time series but has a structure like that of repeated cross-sectional surveys. Moreover, the medians (black lines) give an idea of the trend of prices over time. 2003 has been the most unsatisfying year but also the one with the highest number of sold artworks. After this period, the market experienced a gradual increase in prices and overall turnover. The fall in turnover in 2009, instead, is likely due to the decrease of the number of auctioned items.  However, although the object supply has become scarcer in recent years (compare the low number of sold items in the boxplots and the quite high percentage of sales of Figure~\ref{fig:series}(right)), the turnover is not suffering the same decline due to higher prices. Overall, the positive trend gives an idea of the great potential of the Tribal art market.
\par
In this paper we study the dependence of prices of artworks on available characteristics over the time span 1998-2011 for an overall 14206 items. All hammer prices have been deflated through the HICP (Harmonized Index of Consumer Prices) and transformed in euro. The characteristics of items used as explanatory variables are listed in Table \ref{tab:data1}. Also, based on theoretical arguments we include the interactions of the pairs ``illustration type''--``width of the illustration'' and ``auction house''--``venue''. Indeed, the Cramer's $V$ pairwise association index for such variables is quite high (0.71 and 0.69 respectively). For further details and descriptive analysis of the dataset see~\citet{Candela12} and~\citet{Modugno}.

\begin{table}
  \tiny
     \caption{\label{tab:data1} Covariates of the models.}
    \centering
    \fbox{
    \begin{tabular}{l| l l}
    & \textbf{Variable}&\textbf{Categories}\\
      \hline
\multirow{12}{*}{\emph{Physical}}     & Type of object&  Furniture, Sticks, Masks,\\
        &  &                 Religious objects, Ornaments,\\
        &             &  Sculptures, Musical instruments,\\
         &           &   Tools, Clothing, Textiles,\\
          &          &   Weapons, Jewels \\
    &  Material & Ivory, Vegetable fibre, Wood,\\
    &  &                 Metal, Gold, Stone,\\
    &  &                 Precious stone, Terracotta, ceramic,\\
    &                  & Silver, Textile and hides\\
    &                   &Seashell, Bone, horn, Not indicated  \\
    &  Patina & Not indicated, Pejorative,\\
    &  & Present, Appreciative \\
      \hline
\multirow{19}{*}{\emph{Hystorical}}    &  Region &Central, Southern, Western, Eastern and\\
& &                                               Northern Africa, Australia, Indonesia,\\
& &                                                Melanesia, Polynesia, Mesoamerica, \\
& &                                               Northern and Southern America, Micronesia,\\
& &                                                Far Eastern, Indian Region,  \\
& &                                                Southeastern Asia, Middle East\\
    &  Illustration on the catalogue & Absent, Black/white, Coloured,\\
    &  Illustration width  & Absent, Miscellaneous, \\
       &&                          Quarter page, Half page,\\
     &                           & Full page, More than one,\\
      &                          & Cover\\
  & Description  & Absent, Short visual, Visual,\\
      & &    Broad visual, Critical,       Broad critical.\\
     & Specialized bibliography  &Yes, No\\
     & Comparative bibliography &Yes, No\\
     & Exhibition &Yes, No\\
     &  Historicization  & Absent, Museum certification,\\
     &           & Relevant museum certification,\\
     &           & Simple certification   \\
       \hline
\multirow{2}{*}{\emph{Market}}      &  Venue  & New York, Paris,\\
    &  Auction house  & Sotheby's,            Christie's,\\
    \end{tabular}}
\end{table}

\section{A multilevel model for Tribal art prices}\label{application}
Among the existing proposals for modelling art prices, the hedonic regression model is suitable for Tribal art data. Indeed, such approach seems more suitable for Ethnic art than for Western art data. One reason for this is that Tribal art is considered an \emph{anonymous art} since ethnic objects are not characterized by their artist's name (unknown) but by their ethnic provenance. Since the number of ethnic groups is generally smaller than the number of artists' names, the hedonic model for the Tribal art results in less dummy variables than those applied to other art segments. Moreover, the amount of iconographic subjects and materials is more limited. Therefore, some of the drawbacks of the hedonic regression method are less pronounced when applied to Tribal data. The regression model for the price of artworks corresponding to the hedonic regression specification, that we call ``FE'' (standing for Fixed Effects), can be expressed as

\begin{equation}\label{eq:hedonic}
\log_{10}(y_{it})= \beta_{0t}+ \mathbf{x}^{\T}_{it}\boldsymbol{\beta}+ \epsilon_{it},  \quad \epsilon_{it}|\mathbf{x}^{\T}_{it} \sim \mbox{NID}(0,\sigma^2)
\end{equation}
where $y_{it}$ is the observed price for the time-point $t=1,\ldots,T$ and the item $i=1,\ldots,n_t$ and $\mathbf{x}_{it}$ the correspondent set of covariates listed in Table~\ref{tab:data1}. $\beta_{0t}$ represents the mean price of the time-points $t$. In our specific case, we have chosen to take the semesters as time-points rather than the auction dates, mainly due to three reasons:
 \begin{enumerate}
   \item the auctions are organized in two sessions, one during the winter and one during the summer, and each session contains two to four auctions quite close in time; the concentration in time and space allows to exploit scale economies;
   \item in general, the stakeholders look at the performance of the previous semester;
   \item auction dates are not equally spaced in time and this feature is important for modelling time dependence.
\end{enumerate}
In the dataset, the number of semesters is $T=27$, and  $n_t$, the number of items sold in the semester $t$, varies between 80 (semester 2010-2) and 915 (semester 1998-2); the overall sample size of sold items  (\mbox{$n=\sum_{t=1}^T n_t$}) is 14206.
\par
The FE model fails to capture some essential features of the price dynamics. First, such model is not parsimonious in that both time-effects and categorical covariates are included as dummy variables. Also, the time dummy approach does not allow to model directly the dynamics of prices over time and all the effects are assumed constant over time~\citep{Collins}. Furthermore, potential sources of heterogeneity and heteroscedasticity cannot be accounted for by the hedonic regression model. Last but not least, Tribal art data possess a hierarchical structure which is completely disregarded. For these reasons we propose a multilevel specification which is capable of addressing the aforementioned issues. The task requires a suitable modification of the classic multilevel model. As already highlighted, since we observe different artworks sold at every auction, Tribal art data do not constitute either a panel or a time series. Rather, they can be thought to have a two-level structure in that items, level-1 units, are grouped in time points, level-2 units. Hence, the idea is to exploit the multilevel model to explain heterogeneity of prices among time points.
\par
The two-level model, that we call ``RE'' (standing for Random Effects), has Eq.~(\ref{eq:hedonic}) as the level-1 model, whereas the level-2 model is
\begin{equation}\label{eq:mixed}
\beta_{0t}=\beta_0 + u_t, \quad u_t|\X_t \sim
\mbox{NID}(0,\sigma^2_u), \quad u_t \bot \epsilon_{it}
\end{equation}
where $\beta_0$ is the overall mean price and $u_t$ is a random intercept for the semester $t$. Note that $y_{it}$ and $y_{i(t+1)}$ do not represent the price of item $i$ observed at successive time points, rather, $y_{it}$ indicates the price of the $i$-th object observed at time-point $t$, whereas $y_{i(t+1)}$ is the price of the $i$-th object at time-point $t+1$. The two objects are physically different. One could even specify the temporal dependence of the subscript $i$ by changing it in $i_t$. However, as this would lead to an unnecessary complication in the notation we have chosen the present form.
\par
In the first and in the second column of Table~\ref{tab:results} we present the results of the hedonic regression fit (FE) and the multilevel model (RE) respectively, both of which have been fit through the maximum likelihood method to allow comparisons. The current specification has been driven by both theoretical (Art Economics) and empirical arguments. In practice, all the parameters result significant. Also, notice the magnitude of the effects (with interaction) related to market characteristics such as auction house, venue and illustration.
\par
The two models produce similar results. In particular, besides the estimated coefficients, also the time effects (Semester effect) are very close, although in the FE model these values are estimated coefficients $\hat{\beta}_{0t}$ whereas in the RE model they are Best Linear Unbiased Predictions $\hat{\beta}_0+\hat{u}_t$~\citep{Searle}. This is due to the very high \emph{shrinkage factor}~\citep{Goldstein2010}. In fact, if we consider the cluster means of model~(\ref{eq:hedonic}):
 \begin{equation}\bar{y}_t=\beta_{0t} + \bar{\mathbf{x}}^{\T}_t\boldsymbol{\beta}+\bar{\epsilon}_t,\end{equation}
we have that the estimates of time-specific intercepts correspond to the group means \begin{equation}\hat{\beta}_{0t}=\bar{y}_t-\bar{\mathbf{x}}^{\T}_t\hat{\boldsymbol{\beta}}.\end{equation} On the other hand, the group means for the RE model are obtained as
\begin{equation}
\hat{\beta}_{0t}=\hat{\beta}_0 + \hat{u}_t=\hat{\beta}_0 + \hat{\lambda}_t(\bar{y}_t-\hat{\beta}_0-\bar{\mathbf{x}}^{\T}_t\hat{\boldsymbol{\beta}}),
\end{equation}
where  \begin{equation}\hat{\lambda}_t=\frac{n_t\hat{\sigma}^2_u}{\hat{\sigma}^2+n_t\hat{\sigma}^2_u}\end{equation} is the \emph{shrinkage factor} that can be interpreted as the estimated reliability of the mean raw residual as a predictor of $u_t$. Indeed, the \emph{shrinkage factor} takes values in $[0,1]$ and pulls the group means towards the overall mean by an amount depending both on $n_t$ and on the variance components. Since, in our case, the group sample sizes are big as compared to the variance components, the \emph{shrinkage factor} is close to one for each $t$. Therefore, the time-effects are almost coincident for the two models because each group-specific mean dominates over the population mean.
\par
 Besides the similar parameter estimates, the multilevel model includes a further variability component, the between-group variance, $\sigma^2_u$. The significance of its estimate has been positively assessed through a likelihood ratio test between this model and its unrestricted version (Eq.~(\ref{eq:hedonic}) with $\beta_{0t}=\beta_0$); since the null hypothesis of zero variance is on the boundary of the feasible parameter space, we used half of the $p$-value obtained from the tables of the chi-squared distribution~\citep{Self}. The proportion of the total variability of prices explained by the variability among semesters results \mbox{$100*\sigma^2_u/(\sigma^2_u+\sigma^2)=14.5\%$}, that, in a two-level random-intercept model, corresponds to the \emph{Intra-class correlation} (ICC), the correlation between two observations in the same semester. The existence of a non-zero ICC reveals the inadequacy of traditional modelling frameworks~\citep{Goldstein2010}.
\par
As concerns the diagnostic analysis, the Shapiro-Wilk test points to a deviation from normality in level-1 residuals whereas it does not reject the assumption of normality for level-2 residuals. Given the non-normality at level-1, in order to test the assumption of homogeneity of the variance across clusters, we use a non-parametric version of the homogeneity test of \citet{Levene} which is rank-based \citep{Kruskal}. The results indicate that level-1 variances change over time. To cope with these problems, we have computed robust standard errors for the estimates through a modified version of the \emph{Wild Bootstrap} procedure, described in  subsection~\ref{subsec:wild}. Such scheme is robust with respect to heteroscedastic and non Gaussian errors.
\par
\begin{figure}  \centering
  \includegraphics[width=6cm,height=6cm]{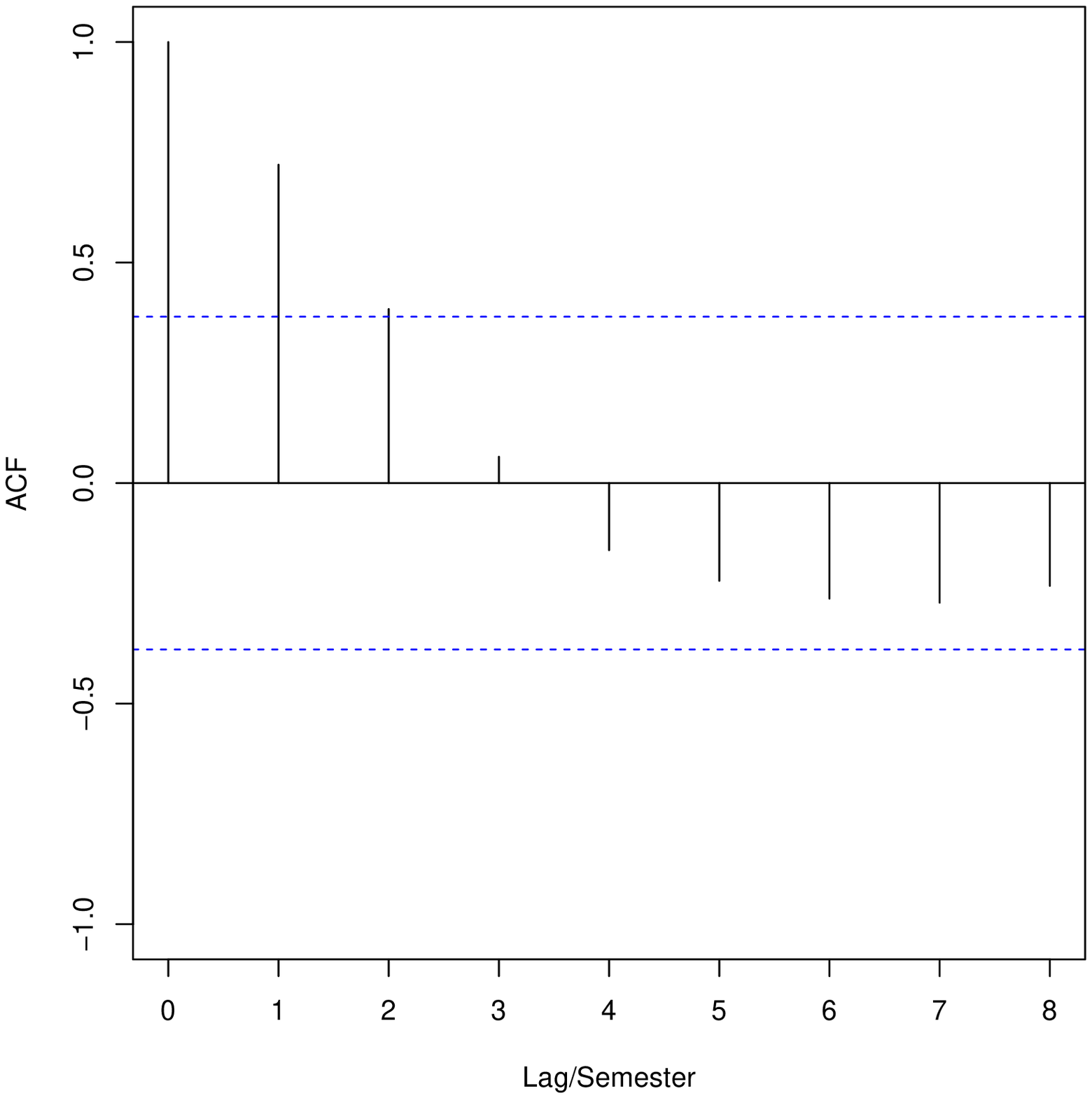}
  \includegraphics[width=6cm,height=6cm]{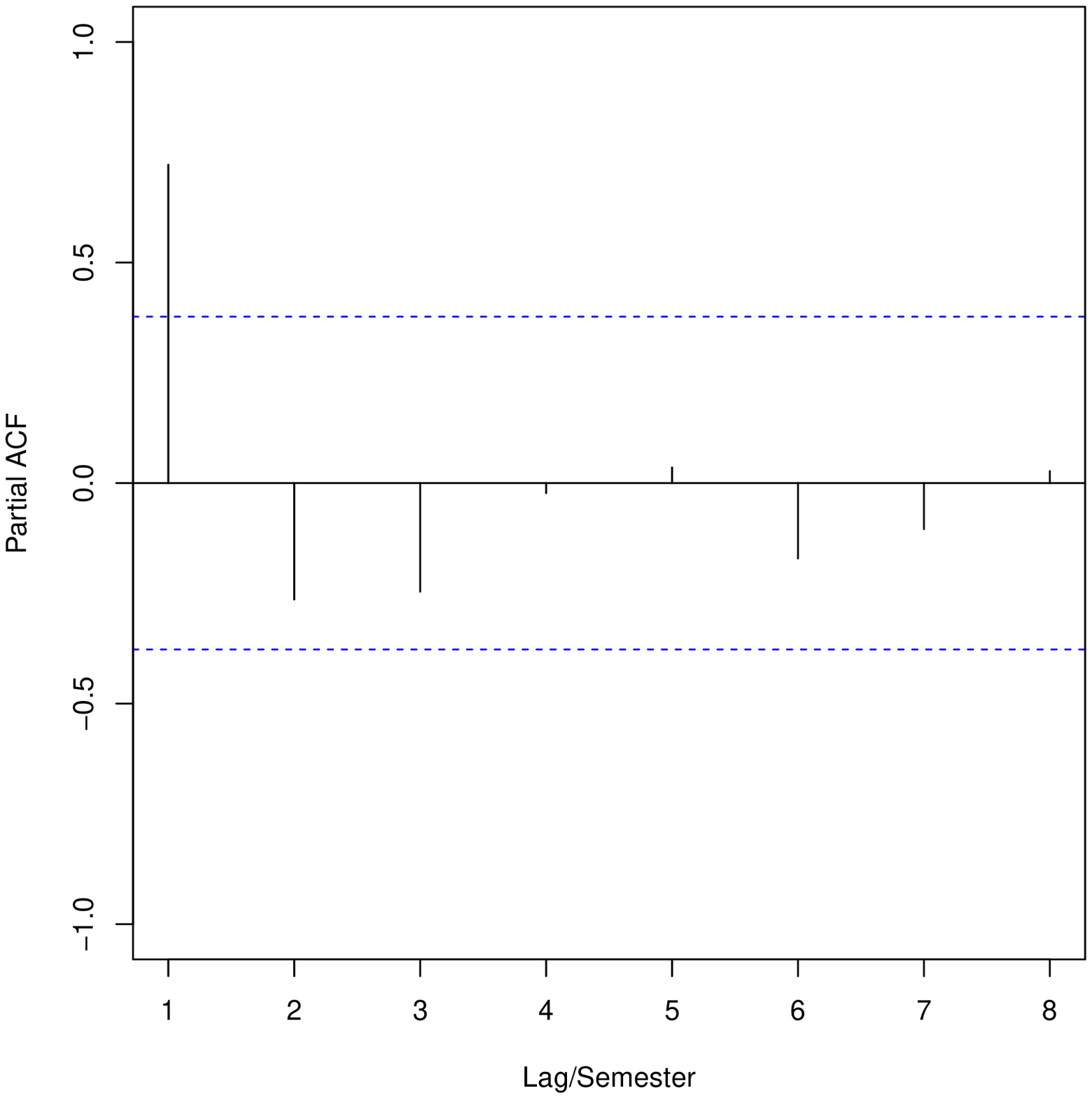}
  \caption{Global (left) and partial (right) autocorrelation functions of level-2 residuals of the RE model.}\label{fig:acf_ind}
\end{figure}

In order to assess the assumption that the error process $u_t$ is a white noise (conditionally to the covariates), we have computed the global and partial autocorrelation functions of level-2 residuals (Figure~\ref{fig:acf_ind}). Clearly, the correlograms point to an autoregressive-like structure, similar to that of an AR(1) process.
In summary, the RE model (\ref{eq:hedonic}) and (\ref{eq:mixed}) produces results very similar to those of the traditional FE model~(\ref{eq:hedonic}) in terms of estimates and residuals, but with greater parsimony. In addition, the multilevel model is able to explain a proportion of variability of the price through the variability among semesters. The assumptions of normality and homogeneity of variance across groups for level-1 errors of both models are not satisfied so that we have used robust bootstrap standard errors. On the other hand, the predicted random effects are normally distributed with zero mean, but they are not independent for different groups as they show a peculiar autocorrelation structure. Improving the classical multilevel model to deal with the latter issue requires relaxing the assumption of independence among random effects. Since in the analysis of Tribal art data these represent time effects, the inclusion of such correlations implies treating them as a time series. As mentioned above, the correlograms of the
residuals suggest the specification of an AR(1) model. Section \ref{newmodel} is devoted to the specification and the estimation of such model.

\subsection{Robust standard errors through the wild bootstrap procedure}\label{subsec:wild}
The \emph{wild bootstrap} was developed by \citet{Liu} following suggestions in \citet{Wu} and \citet{Beran}. Further evidences and refinements for classic regression models are provided in \citet{Flachaire} and \citet{Davidson}. Here, we adopt the wild bootstrap procedure adapted to the case of hierarchical data in \citet{Modugno_boot}.

Consider the random-intercept model for the ($n_t\times 1$) response of the generic group $t$:
\[\y_t=\beta_0+\X_t\boldsymbol{\beta} + \boldsymbol{\nu}_t,\]
where \[\boldsymbol{\nu}_t=\uno_{n_t}u_t+\boldsymbol{\epsilon}_t,\] for all $t=1,\ldots,T$. The disturbances are assumed to be mutually independent and to have zero expectation, but they are allowed to be heteroscedastic. Moreover, the covariates are assumed to be strictly exogenous.\\ Denoting with $\mathbf{H}_t=\X_t(\X^{\T}\X)^{-1}\X_t^{T}$ the orthogonal projection matrix corresponding to design matrix $\X_t$, we replace the residual vector
$\hat{\mathbf{v}}_t= \y_t - \hat{\beta}_0 -  \X_t\boldsymbol{\hat{\beta}}$ by the vector
\[\tilde{\mathbf{v}}_t=\text{diag}\big(\mathbf{I}_{n_t}-\mathbf{H}_t\big) \circ \hat{\mathbf{v}}_t,\]
where the operator ``$\circ$'' denotes the Hadamard (or entrywise) product. Then, the bootstrap procedure used is as follows:
 \begin{enumerate}
 \item draw independently $T$ values, $w_t$, for $t=1,\ldots,T$, from the following two-point auxiliary distribution:
     \begin{equation}\label{eq:F1}
\left\{\begin{array}{lll}
                 -(\sqrt{5}-1)/2 & \text{with probability} & p= (\sqrt{5}+1)/(2\sqrt{5})\\
                 (\sqrt{5}+1)/2 &  \text{with probability} & 1-p,
                 \end{array}\right.
\end{equation}
 with zero mean and unitary variance;
 \item generate the bootstrap samples as \[\y_{t}^*=\hat{\beta}_0+ \X_t\hat{\boldsymbol{\beta}} + \mathbf{\tilde{v}}^*_t \]
     where $\mathbf{\tilde{v}}^*_t=\mathbf{\tilde{v}}_tw_t$,
   \item compute estimates on the bootstrap sample $\y^*$;
   \item repeat steps 1-3 B times and compute bootstrap standard errors as \[\sqrt{\frac{1}{B-1}\sum_{b=1}^B(\boldsymbol{\theta}^*_b-\hat{\boldsymbol{\theta}})^2}\]
where $\hat{\boldsymbol{\theta}}$ is the vector of the ML estimates.
 \end{enumerate}

\citet{Modugno_boot} show that this version of the wild bootstrap behave well in case of heteroscedasticity and non-normality, and, most of all, outperforms the other bootstrap schemes used for multilevel data.
\section{A multilevel model with autoregressive components}\label{newmodel}
In this section we propose an extension of the multilevel model, proposed in Section \ref{application}, that consists in relaxing the assumption of independence among random effects and treating them as a time series at the second level. The section has two subsections: the first one  describes the specification of the model whereas the second subsection presents the implementation of the estimators in the maximum likelihood framework.

\subsection{Model specification}

Consider a  random intercept model with $k$ level-1 covariates: \begin{equation} \label{eq_random_interc}
y_{it}= \beta_{0t}+ \mathbf{x}^{\T}_{it}\boldsymbol{\beta}+ \epsilon_{it},\quad \epsilon_{it}|\mathbf{x}^{\T}_{it} \sim \mbox{NID}(0,\sigma^2) \end{equation}
for $i=1,\ldots,n_{t}$  and $t=1,\ldots,T$. The slopes $\boldsymbol{\beta}$ are fixed;  the intercepts $\beta_{0t}$ are group-specific and random, and they are modeled as \begin{equation}\label{eq:beta0}\beta_{0t}=\beta_0+u_t,\end{equation} where $u_t$ represents the deviation of the group-specific intercept $\beta_{0t}$ from the overall mean, $\beta_0$. The usual assumption of independence for the random effects in~(\ref{eq:mixed})  is relaxed by assuming an autoregressive process of order 1 for level-2 errors:
\begin{equation}\label{eq:AR}
u_t=\rho u_{t-1}+\eta_t, \qquad \eta_t|\X_t \sim \text{NID}(0,\sigma^2_{\eta}), \end{equation}
with $|\rho| < 1$ (that guarantees stationarity), $\eta_t \bot u_s$  and $\eta_t \bot \epsilon_{it}$ for all $s<t$ and for all $i$.

Under these assumptions the dependent variable has the following distribution
\begin{equation}y_{it} \sim N\bigg(\beta_0+ \mathbf{x}_{it}\boldsymbol{\beta},\sigma^2+\phi_0\bigg)\end{equation}
with
\begin{equation}\phi_0=\var(u_t)=\frac{\sigma^2_\eta}{1-\rho^2}.\end{equation}
In matrix form, the composite model for the whole response vector is
\begin{equation}\y = \X\boldsymbol{\beta}+  \Z\mathbf{b} + \boldsymbol{\epsilon},\end{equation}
where $\mathbf{Z}$ is known as random effect design matrix and $\mathbf{b}=\left(\begin{array}{cccc}\beta_{01}&\beta_{02}&\ldots&\beta_{0T}
\end{array}\right)^{\T}$ is the correspondent vector of random intercepts with covariance matrix
\begin{equation}\mathbf{\Gamma}=\phi_0 \left[\begin{array}{cccc}1 & \rho &\ldots &\rho^{T-1}\\
            \rho & 1 &\ldots &\rho^{T-2}\\
            \vdots  & \vdots & \ddots & \vdots \\
             \rho^{T-1} & \rho^{T-2} &\ldots &1\\
\end{array}\right].\end{equation}
Thus, we have:
\begin{equation}\label{eq:y}
\y \sim \text{N}\big(\beta_0+ \X\boldsymbol{\beta}, \Z\mathbf{\Gamma}\Z^{\T}+\sigma^2\mathbf{I}_n \big).
\end{equation}

\subsection{Model estimation}\label{sec:ML}
Model estimation is performed by using the full maximum likelihood estimation method through the E-M algorithm, since the random effects are unobserved.

The set of parameters of the multilevel model with AR(1) random effects to be estimated is $\boldsymbol{\theta}=\{\beta_0,\boldsymbol{\beta},\sigma^2,\rho,\sigma^2_{\eta}\}$. The log-likelihood function associated with the response vector $\y$ is given by
\begin{equation}\label{eq:logfulllik}
\begin{split}
\ell(\boldsymbol{\theta};\mathbf{y}) = \ln \mbox{L}(\boldsymbol{\theta};\mathbf{y})=-\frac{n}{2}\ln (2\pi)-\frac{1}{2}\ln |\mathbf{\Omega}| - \frac{1}{2} (\y-\dot{\X}\dot{\boldsymbol{\beta}})^{\T}\mathbf{\Omega}^{-1}(\y-\dot{\X}\dot{\boldsymbol{\beta}}).
\end{split}
\end{equation}
where
\begin{equation}\mathbf{\Omega}=\Z\mathbf{\Gamma}\Z^{\T}+\sigma^2\mathbf{I}_n\end{equation}
is the covariance matrix of $\y$, and
\begin{equation}\dot{\X}=\left[\begin{array}{cc}\uno_n & \X\end{array}\right] \qquad \text{and} \qquad \dot{\boldsymbol{\beta}}=\left[\begin{array}{cc}\beta_0 & \boldsymbol{\beta}\end{array}\right]^{\T}\end{equation}
are the matrix design and the coefficients vector including the intercept, respectively.

To simplify the notation, we separate the set of parameters of the model into two subsets: $\boldsymbol{\theta}=\{\boldsymbol{\theta}_1,\boldsymbol{\theta}_2\}$, where the subset $\boldsymbol{\theta}_1=\{\boldsymbol{\beta},\sigma^2\}$ includes the level-1 parameters, and $\boldsymbol{\theta}_2=\{\beta_0,\rho,\sigma^2_{\eta}\}$ is the set of level-2 parameters.

The complete log-likelihood of the observed and unobserved data can be expressed as the sum of two separate components
 \begin{equation}\label{loglik}
\begin{split}
\ell(\boldsymbol{\theta};\mathbf{y},\mathbf{b}) = \ln \mbox{L}(\boldsymbol{\theta};\mathbf{y},\mathbf{b})= \ell_1(\boldsymbol{\theta}_1)+\ell_2(\boldsymbol{\theta}_2)
\end{split}
\end{equation}
where
\begin{equation*}
\ell_1(\boldsymbol{\theta}_1)=\ln f(\y|\mathbf{b})=-\frac{n}{2}\ln (2\pi\sigma^2) -\frac{(\mathbf{y}-\X\boldsymbol{\beta}-\Z\mathbf{b})^{\T}(\mathbf{y}-\X\boldsymbol{\beta}-\Z\mathbf{b})}{2\sigma^2}
\end{equation*}
and
\begin{equation}
\label{eq:l2}\ell_2(\boldsymbol\theta_2)=\ln f(\mathbf{b})=-\frac{T}{2} \ln (2\pi\sigma^2_{\eta})+\frac{1}{2}\ln (1-\rho^2) -\frac{(\mathbf{b}-\beta_0\uno_T)^{\T}\mathbf{V}^{-1}(\mathbf{b}-\beta_0\uno_T)}{2\sigma^2_{\eta}}.
\end{equation}

The matrix $\V=\frac{1}{\sigma^{2}_{\eta}}\boldsymbol\Gamma$ and it is straightforward to show that~\citep{Hamilton} \begin{equation}\V^{-1}=\left[\begin{array}{cccccc}
1 & -\rho & 0 &\ldots & 0 & 0\\
-\rho & 1+\rho^2 & -\rho& \ldots&0&0\\
0 & -\rho & 1+\rho^2 & \ldots & 0 & 0\\
\vdots & & & &\vdots\\
0 & 0 & 0 & \ldots & 1+\rho^2 & -\rho\\
0 & \ldots & & &-\rho & 1 \end{array}\right].\end{equation}
The estimation of $\boldsymbol{\theta}$ through the E-M algorithm  consists of two steps, the Expectation (E) and Maximization (M) step described in detail in the following.
\begin{description}
  \item[\textbf{E step}]  In the expectation step the expected score functions of the parameter conditioned to the observed data are computed on the basis of current value of $\boldsymbol\theta$, denoted as $\boldsymbol\theta^{h}$, as follows:
  \begin{equation}\label{modifloglik}
  \E\big[S(\boldsymbol{\theta};\mathbf{y},\mathbf{b})|\y,\boldsymbol{\theta}^{(h)}\big]=\E\big[S_{1}(\boldsymbol{\theta}_{1})|\y,\boldsymbol{\theta}^{(h)}\big]+\E\big[S_{2}(\boldsymbol{\theta} _{2})|\y,\boldsymbol{\theta}^{(h)}\big]
  \end{equation}
  where $S(\boldsymbol{\theta})=\partial \ell(\boldsymbol{\theta};\mathbf{y})/\partial\boldsymbol{\theta}$, $S_{1}(\boldsymbol{\theta}_{1})=\partial \ell_1(\boldsymbol{\theta}_{1})/\partial\boldsymbol{\theta}_{1}$ and $S_{2}(\boldsymbol{\theta}_{2})=\partial \ell_2(\boldsymbol{\theta}_{2})/\partial\boldsymbol{\theta}_{2}$.\\
  The expressions of the expected score functions with respect to level-1 parameters of the model are given by
  \begin{eqnarray}\label{eq:exp_score1}
  \E({ S_1(\boldsymbol{\beta})}|\y;\boldsymbol{\theta}^{(h)})&=&\frac{1}{\sigma^2}(-\X^{\T}\X^{\T}\boldsymbol{\beta}+\X^{\T}\y-\X^{\T}\Z \hat{\mathbf{b}})\\\nonumber
  \E({ S_1(\sigma^2)}|\y;\boldsymbol{\theta}^{(h)})&=&-\frac{n}{2\sigma^2}+\frac{(\y-\X\boldsymbol{\beta} - \Z \hat{\mathbf{b}})^{\T}(\y-\X\boldsymbol{\beta} - \Z \hat{\mathbf{b}})+\mbox{tr}(\Z^{\T}\Z \mathbf{B})}{2\sigma^4}.\\ \nonumber
  \end{eqnarray}
  The expression of the expected score functions with respect to the level-2 parameters of the model are
  \begin{eqnarray*}\label{eq:exp_score2}
  \E({ S_2(\beta_0)}|\y;\boldsymbol{\theta}^{(h)})&=&\frac{(1-\rho)(\hat{b}_1+\hat{b}_T)+(1-\rho)^2\sum_{t=2}^{T-1}\hat{b}_t - (1-\rho)(T-(T-2)\rho)\beta_0}{\sigma^2_{\eta}}\\ \nonumber
  \E({ S_2(\sigma^2_{\eta})}|\y;\boldsymbol{\theta}^{(h)})&=&-\frac{T}{2\sigma^2_{\eta}}+\frac{\mbox{tr}(\mathbf{V}^{-1} \mathbf{B})+\hat{\ub}^{\T}\mathbf{V}^{-1} \hat{\ub}}{2\sigma^4_{\eta}}\\ \nonumber
  \E(S_2(\rho)|\y;\boldsymbol{\theta}^{(h)})&=& \frac{1}{\sigma^2_{\eta}}\Bigg[ \sum_{t=1}^{T-1}(\mathbf{B}_{t,t+1}+\hat{u}_t\hat{u}_{t+1})-\rho \sum_{t=2}^{T-1}(\mathbf{B}_{t,t}+\hat{u}_t^2)\Bigg] - \frac{\rho}{1-\rho^2}
  \end{eqnarray*}
  where \\
  \begin{eqnarray}\label{eq:posterior}
  \hat{\ub}&=& \E(\mathbf{u}|\y)=\mathbf{\Gamma}\Z^{\T}\mathbf{\Omega}^{-1}(\y-\uno_n\beta_0 - \X\boldsymbol{\beta}) \\
  \hat{\mathbf{b}}&=&\hat{\beta}_0+\hat{\ub}\\
  \mathbf{B}&=&\var(\mathbf{b}|\y)=\mathbf{\Gamma}-\mathbf{\Gamma}\Z^{\T}\mathbf{\Omega}^{-1}\Z\mathbf{\Gamma}^{\T} \nonumber
  \end{eqnarray}

  \item[\textbf{M step}] It consists in maximizing the conditional expected value of the log-likelihood~(\ref{modifloglik}) computed in the E-step, getting maximum likelihood estimates of the model parameters. In detail, the current values of vector of parameters $\boldsymbol{\theta}^{(h+1)}$ are updated as follows
  \begin{eqnarray*}
  \hat{\boldsymbol{\beta}}^{(h+1)}&=&(\X^{\T}\X)^{-1}\X^{\T}(\y-\Z\hat{\mathbf{b}})\\
  (\hat{\sigma}^2)^{(h+1)}&=&\frac{(\y-\X\boldsymbol{\beta} - \Z \hat{\mathbf{b}})^{\T}(\y-\X\boldsymbol{\beta} - \Z \hat{\mathbf{b}})+\mbox{tr}(\Z^{\T}\Z \mathbf{B})}{n}\\
  \hat{\beta}_0^{(h+1)}&=&\frac{b_1+b_T+(1-\rho)\sum_{t=2}^{T-1}b_t}{T-(T-2)\rho}\\
  (\hat{\sigma}^2_{\eta})^{(h+1)}&=&\frac{\mbox{tr}(\mathbf{V}^{-1} \mathbf{B})+\hat{\ub}^{\T}\mathbf{V}^{-1} \hat{\ub}}{T}
  \end{eqnarray*}
  Since we get non linear maximum likelihood equation for the parameter $\rho$,  we update its current value through an iteration of the Newton-Raphson scheme.\\
\end{description}

The E-M algorithm consists of the following steps
\begin{enumerate}
  \item Choose an initial value for the parameters $\boldsymbol{\theta}$;
  \item Compute the expected score functions for all the parameters (E-step);
  \item Obtain improved parameter estimates (M-step);
  \item Repeat steps 2 and 3 until convergence, that is, until \begin{equation}\ell(\boldsymbol{\theta}^{(h+1)};\mathbf{y},\mathbf{b})-\ell(\boldsymbol{\theta}^{(h)};\mathbf{y},\mathbf{b})\end{equation} is arbitrarily small.
\end{enumerate}

The EM algorithm produces the Empirical Bayes prediction for the random effects $\mathbf{b}$, namely, the mean of their conditional distribution with respect to the observed data $\y$ as in~(\ref{eq:posterior})~\citep{Searle}. The whole algorithm has been implemented in R with an original code. Further details on the implementation and a Monte Carlo study based on the code can be found in \citet{modugnoPhD}.

\section{Application of the new model to Tribal art data}\label{app}
In this section, we present the results of the fit of the new model upon the Tribal art dataset. Moreover, we will compare the predicting capability of the three models under scrutiny. Consider the model in equations~(\ref{eq_random_interc}), (\ref{eq:beta0}) and~(\ref{eq:AR}), with the same set of covariates as in the FE specification reported in Table~\ref{tab:data1}. We call it ``ARE'' standing for Autoregressive Random Effects. The results are shown in the third column of Table~\ref{tab:results}. The estimates and the predicted random effects are quite close to those from the RE model (second column). In this case, the estimated between-group variance, that takes the form $\text{Var}(u_t)=\sigma^2_{\eta}/(1-\rho^2)$, results $0.036$, slightly bigger than that of the RE model ($\hat{\sigma}^2_u=0.029$). Consequently, the proportion of variability explained by the between-semesters variance (ICC) is bigger for the new model, 17.3\% against 14.5\%. Also, the level-2 residual variability of the ARE model $\hat{\sigma}^2_{\
eta}=0.010$ is smaller than that of the RE model, $\hat{\sigma}^2_u=0.029$.  This confirms that the structure at the second level has been taken into account by the new specification. Furthermore, the estimate of the autoregressive parameter $\rho$ is quite high, $\hat{\rho}=0.843$ and agrees with the evidence of the correlograms of the residuals of the RE model (see Figure~\ref{fig:acf_ind}). The last column reports $\hat{\beta}_0+\hat{u}_t$ of the ARE model to facilitate the comparison with the FE semester effects.
\par
Note that when $\rho$ is zero, the random effects are independent and the multilevel model reduces to the RE specification. Hence, the ARE and the RE models are nested so that we can use the likelihood ratio test for assessing the significance of $\rho$. According both to the LR test and to the Information Criteria (see Table~\ref{tab:results}), the ARE model provides a better fit than the RE model.
\begin{table}
    \tiny
    \caption{\label{tab:results}Results of the fit for models FE (\ref{eq:hedonic}), RE (\ref{eq:hedonic}) and (\ref{eq:mixed}) and ARE (\ref{eq_random_interc}), (\ref{eq:beta0}) and (\ref{eq:AR}). Bootstrap standard errors in parentheses.}
    \centering
\fbox{    \begin{tabular}{lcccc}
    \hline
     & FE & RE &  \multicolumn{2}{c}{ARE} \\
  \hline
    AIC & 15576  & 15671  & 15647  &\\
  BIC & 16317  & 16223  & 16207  &\\
  \# param. & 98  & 73  & 74  &\\
\hline
  $\hat{\sigma}^2$ & 0.173 (0) & 0.173 (0.036) & 0.173 (0.043) &\\
  $\hat{\sigma}^2_u$ & -  & 0.029 (0.009) & - &\\
  $\hat{\sigma}^2_{\eta}$ - & - & 0.01 (0.013)& \\
  $\hat{\rho}$ &  - & - & 0.843 (0.128) &\\
  ICC      &   -  & 0.145  & 0.173  &\\
  \hline
   $\hat{\beta}_0$         & -            & 2.216 (0.068) & 2.212 (0.112) &\\
      \hline
     Semester effect   &$\hat{\beta}_{0t}$& $\hat{u}_t$ & $\hat{u}_t$    & $\hat{\beta}_0+\hat{u}_t$\\
  \hline
  1998-1 & 1.96 (0.075) & -0.254 (0.022) & -0.257 (0.021) & 1.983 \\
  1998-2 & 2.081 (0.07) & -0.137 (0.016) & -0.143 (0.022) & 2.097 \\
  1999-1 & 2.15 (0.072) & -0.068 (0.019) & -0.069 (0.025) & 2.170 \\
  1999-2 & 2.355 (0.072) & 0.135 (0.017) & 0.13 (0.023)   & 2.369 \\
  2000-1 & 2.454 (0.071) & 0.234 (0.016) & 0.229 (0.023)  & 2.468 \\
  2000-2 & 2.418 (0.071) & 0.197 (0.016) & 0.195 (0.021)  & 2.435 \\
  2001-1 & 2.393 (0.074) & 0.171 (0.02) & 0.165 (0.025)   & 2.405 \\
  2001-2 & 2.244 (0.077) & 0.025 (0.025) & 0.038 (0.029)  & 2.277 \\
  2002-1 & 2.352 (0.071) & 0.133 (0.017) & 0.12 (0.024)   & 2.360 \\
  2002-2 & 2.15 (0.075) & -0.066 (0.024) & -0.068 (0.028) & 2.171 \\
  2003-1 & 2.031 (0.073) & -0.185 (0.017) & -0.191 (0.023)& 2.048 \\
  2003-2 & 1.932 (0.071) & -0.283 (0.016) & -0.29 (0.023) & 1.949 \\
  2004-1 & 1.911 (0.072) & -0.304 (0.019) & -0.31 (0.025) & 1.930 \\
  2004-2 & 2.029 (0.072) & -0.186 (0.016) & -0.193 (0.024)& 2.047 \\
  2005-1 & 2.204 (0.073) & -0.014 (0.018) & -0.025 (0.025)& 2.215 \\
  2005-2 & 2.175 (0.073) & -0.043 (0.017) & -0.048 (0.024)& 2.191 \\
  2006-1 & 2.192 (0.072) & -0.025 (0.019) & -0.035 (0.025)& 2.205 \\
  2006-2 & 2.09 (0.073) & -0.126 (0.016) & -0.13 (0.023)  & 2.109 \\
  2007-1 & 2.151 (0.072) & -0.066 (0.017) & -0.073 (0.024)& 2.166 \\
  2007-2 & 2.195 (0.07) & -0.023 (0.019) & -0.03 (0.027)  & 2.209 \\
  2008-1 & 2.196 (0.068) & -0.022 (0.022) & -0.031 (0.028)& 2.208 \\
  2008-2 & 2.119 (0.074) & -0.098 (0.017) & -0.101 (0.025)& 2.139 \\
  2009-1 & 2.225 (0.07) & 0.006 (0.019) & 0.003 (0.026)   & 2.243 \\
  2009-2 & 2.438 (0.08) & 0.213 (0.032) & 0.201 (0.037)   & 2.440 \\
  2010-1 & 2.449 (0.079) & 0.226 (0.03) & 0.227 (0.035)   & 2.466 \\
  2010-2 & 2.523 (0.097) & 0.282 (0.056) & 0.285 (0.055)  & 2.525 \\
  2011-1 & 2.506 (0.075) & 0.28 (0.033) & 0.279 (0.03)    & 2.519 \\
\hline
\multicolumn{5}{c}{Type of object: baseline Furniture}\\
\hline
 Sticks & -0.093 (0.026) & -0.093 (0.035) & -0.094 (0.028) &\\
  Masks & 0.109 (0.021) & 0.109 (0.023) & 0.108 (0.024) &\\
  Religious objects & 0.001 (0.023) & 0.001 (0.028) & -0.001 (0.027) &\\
  Ornaments & -0.097 (0.026) & -0.097 (0.036) & -0.099 (0.029) &\\
  Sculptures & 0.049 (0.02) & 0.049 (0.024) & 0.047 (0.023) &\\
  Musical instruments & -0.117 (0.033) & -0.116 (0.045) & -0.118 (0.038)& \\
  Tools & -0.084 (0.021) & -0.084 (0.024) & -0.085 (0.024)& \\
  Clothing & -0.068 (0.039) & -0.068 (0.055) & -0.069 (0.042) &\\
  Textiles & -0.04 (0.038) & -0.04 (0.053) & -0.041 (0.041) &\\
  Weapons & -0.097 (0.027) & -0.097 (0.034) & -0.098 (0.029) &\\
  Jewels & -0.045 (0.034) & -0.045 (0.046) & -0.047 (0.038) &\\
  \hline
\multicolumn{5}{c}{Yes vs No}\\
\hline
Specialized bibliography (dummy)& 0.14 (0.012) & 0.14 (0.02) & 0.14 (0.013) &\\
Comparative bibliography (dummy)& 0.118 (0.009) & 0.119 (0.021) & 0.119 (0.01) &\\
Exhibition (dummy)              & 0.08 (0.014) & 0.08 (0.028) & 0.08 (0.015) &\\
    \hline
  \multicolumn{5}{c}{Historicization: baseline Absent}\\
  \hline
  Museum certification          & 0.009 (0.015) & 0.009 (0.04) & 0.01 (0.017) &\\
  Relevant museum certification & 0.015 (0.016) & 0.015 (0.042) & 0.016 (0.016) &\\
  Simple certification          & 0.032 (0.01) & 0.032 (0.031) & 0.032 (0.01) \\
   \hline
  \multicolumn{5}{r}{\emph{(continued in the next page)}}\\
    \end{tabular}}
\end{table}
\begin{table}
    \tiny
    \addtocounter{table}{-1}
    \caption{\emph{(continued from the previous page)}}
    \centering
 \fbox{   \begin{tabular}{lcccc}
    \hline
     & FE & RE &  \multicolumn{2}{c}{ARE} \\
   \hline
\multicolumn{5}{c}{Region: baseline Central America}\\
\hline
Southern Africa       & -0.164 (0.033) & -0.165 (0.04) & -0.165 (0.036) &\\
 Western Africa       & -0.105 (0.012) & -0.106 (0.017) & -0.106 (0.012) &\\
Eastern Africa       & -0.161 (0.029) & -0.162 (0.035) & -0.162 (0.032) &\\
Australia            &     0.038 (0.053) & 0.038 (0.088) & 0.037 (0.055) &\\
Indonesia            &     -0.111 (0.027) & -0.112 (0.046) & -0.112 (0.029)& \\
Melanesia              &   0.007 (0.016) & 0.007 (0.031) & 0.006 (0.016) &\\
Polynesia                & 0.185 (0.018) & 0.184 (0.032) & 0.184 (0.02) &\\
Northern  America         & 0.232 (0.018) & 0.232 (0.056) & 0.232 (0.02) &\\
Northern Africa           &-0.374 (0.123) & -0.375 (0.18) & -0.375 (0.13) &\\
Southern  America        &     0.013 (0.023) & 0.012 (0.051) & 0.013 (0.025) &\\
Mesoamerica &0.114 (0.021) & 0.113 (0.051) & 0.114 (0.021) &\\
Far Eastern              &         -0.06 (0.139) & -0.061 (0.315) & -0.06 (0.15) &\\
Micronesia                 &    0.097 (0.076) & 0.097 (0.078) & 0.097 (0.08) &\\
Indian Region                       &    0.303 (0.096) & 0.299 (0.097) & 0.296 (0.092)& \\
Asian Southeast                  &    -0.064 (0.118) & -0.066 (0.153) & -0.067 (0.125) &\\
Middle East                      &   -0.514 (0.085) & -0.513 (0.164) & -0.513 (0.088)& \\
    \hline
 \multicolumn{5}{c}{Type of material: baseline Ivory}\\
  \hline
 Vegetable fibre, paper, plumage & -0.046 (0.028) & -0.046 (0.041) & -0.047 (0.032) &\\
  Wood & 0.078 (0.021) & 0.078 (0.029) & 0.077 (0.024) &\\
  Metal & -0.033 (0.028) & -0.034 (0.046) & -0.035 (0.033) &\\
  Gold & 0.13 (0.032) & 0.13 (0.059) & 0.129 (0.037) &\\
  Stone & 0.046 (0.03) & 0.046 (0.036) & 0.045 (0.034)& \\
  Precious stone & 0.052 (0.033) & 0.052 (0.045) & 0.052 (0.037) &\\
  Terracotta, ceramic & 0.007 (0.027) & 0.007 (0.044) & 0.006 (0.031) &\\
  Silver & -0.079 (0.048) & -0.08 (0.078) & -0.08 (0.047) &\\
  Textile and hides & -0.019 (0.033) & -0.019 (0.058) & -0.021 (0.04)& \\
  Seashell & 0.058 (0.054) & 0.058 (0.11) & 0.057 (0.059)& \\
  Bone, horn & -0.13 (0.036) & -0.131 (0.07) & -0.131 (0.039) &\\
  Not indicated & 0.044 (0.045) & 0.044 (0.055) & 0.041 (0.052)& \\
    \hline
  \multicolumn{5}{c}{Patina: baseline Not indicated}\\
  \hline
   Pejorative & 0.235 (0.039) & 0.234 (0.044) & 0.234 (0.04) &\\
   Present & 0.029 (0.011) & 0.028 (0.025) & 0.028 (0.013) &\\
   Appreciative & 0.11 (0.012) & 0.109 (0.029) & 0.109 (0.013) &\\
  \hline
  \multicolumn{5}{c}{Description on the catalogue: baseline Absent}\\
  \hline
   Short visual descr. & -0.13 (0.037) & -0.132 (0.101) & -0.134 (0.038) &\\
   Visual descr. & 0.039 (0.038) & 0.038 (0.103) & 0.036 (0.039)& \\
   Broad visual descr. & 0.279 (0.041) & 0.278 (0.114) & 0.276 (0.043) &\\
   Critical descr. & 0.269 (0.041) & 0.268 (0.119) & 0.266 (0.042)& \\
   Broad critical descr. & 0.634 (0.046) & 0.634 (0.131) & 0.632 (0.048) &\\
\hline
  \multicolumn{5}{c}{Illustration: baseline Absent}\\
\hline
  Miscellaneous col. & 0.411 (0.02) & 0.412 (0.048) & 0.41 (0.021) &\\
  Col. cover & 1.412 (0.11) & 1.411 (0.202) & 1.41 (0.113) &\\
  Col. half page & 0.854 (0.023) & 0.856 (0.072) & 0.854 (0.024) &\\
  Col. full page & 1.008 (0.025) & 1.008 (0.075) & 1.007 (0.024)& \\
  More than one col. & 1.223 (0.028) & 1.223 (0.078) & 1.221 (0.029)& \\
  Col. quarter page & 0.674 (0.021) & 0.675 (0.062) & 0.673 (0.021)& \\
  Miscellaneous b/w & 0.41 (0.033) & 0.409 (0.055) & 0.406 (0.035)& \\
  b/w half page & 0.551 (0.045) & 0.552 (0.084) & 0.549 (0.051)& \\
  b/w quarter page & 0.304 (0.025) & 0.305 (0.075) & 0.303 (0.027) &\\
  \hline
  \multicolumn{5}{c}{Auction house and venue: baseline Bonhams-New York}\\
  \hline
  Christie's-Amsterdam & 0.765 (0.054) & 0.766 (0.073) & 0.756 (0.059) &\\
  Christie's-New York & 0.702 (0.054) & 0.7 (0.055) & 0.69 (0.059)& \\
  Christie's-Paris & 0.601 (0.05) & 0.6 (0.049) & 0.592 (0.054)& \\
  Encheres Rive Gauche-Paris & 0.536 (0.086) & 0.534 (0.044) & 0.523 (0.09)& \\
  Koller-Zurich & -0.012 (0.052) & -0.014 (0.075) & -0.021 (0.059) &\\
  Piasa-Paris & 0.753 (0.071) & 0.751 (0.052) & 0.74 (0.071) &\\
  Sotheby's-New York & 0.866 (0.049) & 0.866 (0.046) & 0.856 (0.055) &\\
  Sotheby's-Paris & 0.761 (0.05) & 0.761 (0.049) & 0.752 (0.055) &\\
      \hline
    \end{tabular}               }
\end{table}

The diagnostic checks show that the ARE-model presents the same features of non-normality and non-homogeneity of variance among groups as the FE-model (section \ref{application}). Therefore, also in this case, we have computed robust standard errors through the wild bootstrap procedure. The presence of AR(1) random effects requires a further extension of the wild bootstrap for hierarchical data \citep{Modugno_boot} that consists in replacing step (b) of subsection \ref{subsec:wild} with the following:
\begin{enumerate}
    \item[(b)] generate the bootstrap samples as
    \[
    y_{it}^*=\hat{\beta}_0+ \mathbf{x}_{it}^{\T}\hat{\boldsymbol{\beta}} + u_t^* + \epsilon_{it}^*
    \]
    for $i =1, \ldots, n$ and for $t=1,\ldots,T$,
    where $u_t^*$ is an autoregressive process with disturbances equal to  $w_{t\eta}\hat{\eta}_t$ and
    \[
    \epsilon_{it}^*=\hat{\epsilon}_{it}/(1-h_i) w_{t\epsilon};
    \]
     $h_i$ is $i$-th diagonal element of the orthogonal projection matrix of $\X$;
\end{enumerate}
 such modification takes into account both the time dependence at the second level and the heteroscedasticity at the first level.
\par
\begin{figure}
  \centering
  \includegraphics[width=6cm,height=6cm]{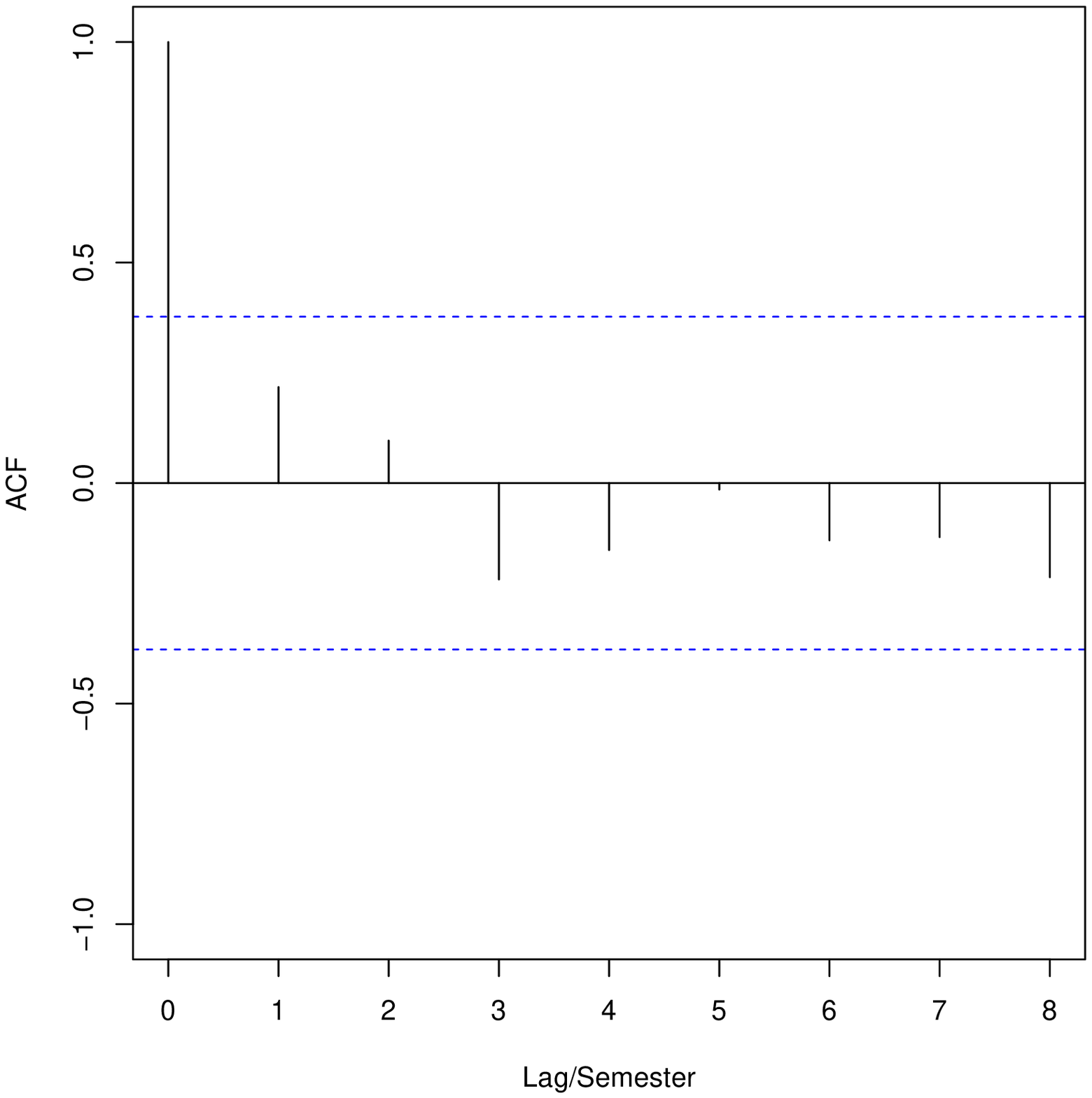}
  \includegraphics[width=6cm,height=6cm]{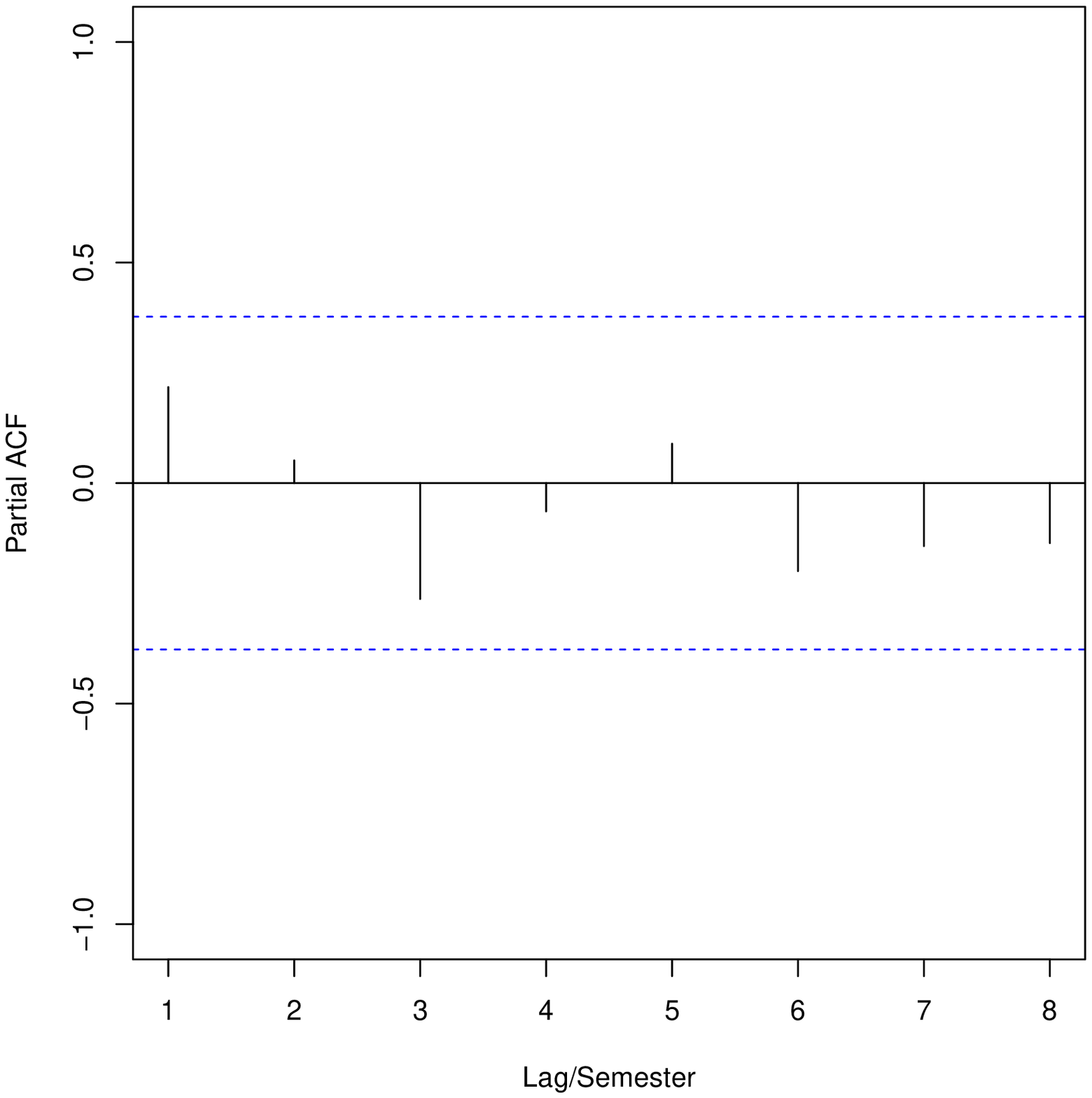}
  \caption{Plots of autocorrelation functions of the level-2 residuals of the ARE model.}\label{fig:acf_ar}
\end{figure}

The autocorrelation functions (global and partial) of level-2 residuals (see Figure~\ref{fig:acf_ar}) do not reveal any structure as the values lie within the rejection bands at level 95\% at all lags. Hence, our novel specification has successfully captured the time dependence of the price dynamics by means of the autoregressive specification at the second level.
Finally, Table~\ref{tab:prediction} summarizes and compares the prediction capability of the three models under study. The aggregate measures of prediction error are the Mean Absolute (Prediction) Error $\text{MAE}=\frac{1}{n_{T+1}}\sum_{i=1}^{n_t}\left|y_{i,T+1}-\hat{y}_{i,T+1}\right|$ and the Root Mean Square (Prediction) Error $\text{RMSE}=\sqrt{\frac{1}{n_{T+1}}\sum_{i=1}^{n_{T+1}}\big(y_{i,T+1}-\hat{y}_{i,T+1}\big)}$. The first two rows of Table~\ref{tab:prediction} report the prediction error over 100 (out of sample) items within the time span 1998-2011. In this instance, the three models present the same performance. The last two rows of the table show the forecast performance over all the 281 observations of the semester 2011-1. Such observations have not been included in the model so that the measures reflect a genuine one-step-ahead forecast performance. Clearly, the ARE model allows to obtain better forecasts of the prices of artwork objects through the autoregressive specification.

\begin{table}
  \small
  \caption{Prediction/forecasting performance of the three models over 100 out-of-sample units within the time span 1998-2011 (rows 1-2) and over 281 units of the out-of-sample semester, $2011-1$ (rows 3-4).}\label{tab:prediction}
    \centering
  \begin{tabular}{lccc}
  \hline
    & \textbf{FE}& \textbf{RE} & \textbf{ARE}\\
    \hline
    \multicolumn{4}{c}{\emph{$100$ units within the time span} 1998-2011}\\
    MAE  & 0.280 & 0.280 & 0.280 \\
    RMSE & 0.342 & 0.342 & 0.342 \\
    \hline
    \multicolumn{4}{c}{\emph{$281$ units in the semester} 2011-1}\\
    MAE  & 0.494 & 0.489 & 0.454 \\
    RMSE & 0.423 & 0.419 & 0.358 \\
    \hline
    \end{tabular}
\end{table}

In conclusion, if compared to the other two models, our new ARE model presents a better fit and superior forecasting performance. Although the estimates are similar to those of the hedonic regression model, the multilevel framework is more parsimonious and provides a natural flexible approach through the decomposition of the total variability of the response. The autoregressive specification is backed up by Art Economics theory that confirms that the process of formation of auction prices has short memory: indeed, in the case of Tribal Art, the dependence is upon the previous semester.

\section{Conclusions}\label{sec:concl}
In the present work, we have introduced a multilevel framework for the analysis of prices of artworks sold at auctions over time. The proposal combines the flexibility of mixed effect models, in that it allows to account for various sources of heterogeneity, together with the predicting performance of time series models. The latter component allows to specify a substantive model for the price dynamics over time. Since auction data do not constitute a proper panel or a time series we need a multilevel specification with items at the first level and time points at the second level.
\par
We have applied such specification to analyse the Tribal art market by using the first database on Ethnic artworks that contains information on more than 20000 items sold in the most important auction houses in the world. The results show that our approach gives a substantial advantage over the traditional hedonic regression model, especially in terms of degrees of freedom, parsimony and interpretability. In fact, the multilevel model retains the ease of interpretation of the hedonic regression model since the estimated regression coefficients can be still seen as \emph{shadow prices} for each feature, and a price index for the art market is easily provided through the predictions of the time-effects. On the other hand, it has less parameters to be estimated and provides a decomposition of the total variability of the response.
\par
The dependence of the price over time has been modelled by means of an autoregressive specification at the second level. Hence, we have extended the classic multilevel model by relaxing the assumption of independence among random effects and treating them as a time series at the second level. In order to achieve the task, we have derived full maximum likelihood estimators through the E-M algorithm and have implemented them in an original R-code. The results show that the new specification fully captures the temporal dependence structure among group-effects. Moreover, such model presents superior forecasting performance with respect to other proposals. In conclusion, we advocate the use of our specification as a natural choice for modelling artwork prices and possibly, obtain forecasts/predictions that might be valuable to auction houses, banks and investors.
\par
The work presented here can be extended in different directions; also, many applications are possible. First, it could be interesting to explore further the nature of the deviation from normality of level-1 residuals. This might be accomplished by inserting further variance components in the model, especially those related to the interactions between covariates. Also, possible volatility effects (ARCH/GARCH) can be inserted as to extend considerably the flexibility of the model and make it appealing from the point of view of financial applications. Moreover, the model could be applied to characterize and forecast other art markets. Lastly, in order to promote the usage of our model and to facilitate the reproducibility of the research we plan to release the software implemented as an R package. The latter project would contribute to fill the lack that hindered the practical use of multilevel models for repeated cross-sectional data.

\section*{Acknowledgements}
We would like to thank Leonardo Grilli, Cinzia Viroli, Guido Candela and Antonello E. Scorcu for their useful comments and discussions. This work has been partially supported by the FIRB Research project ``Mixture and latent variable model for causal inference and analysis of socio-economic data''.

\bibliographystyle{chicago}

\end{document}